\documentclass[12pt]{article}
\usepackage{epsfig}

\usepackage{a4}

\usepackage{amsmath}

\begin{document}
\title{Static BPS 'monopoles' in all even spacetime dimensions}
\author{{\large Eugen Radu}$^{\dagger}$
and {\large D. H. Tchrakian}$^{\dagger \star}$ \\ \\
$^{\dagger}${\small Department of
Mathematical Physics, National University of Ireland Maynooth,} \\
{\small Maynooth, Ireland} \\
$^{\star}${\small School of Theoretical Physics -- DIAS, 10 Burlington
Road, Dublin 4, Ireland }}

\date{}
\newcommand{\dd}{\mbox{d}}
\newcommand{\tr}{\mbox{tr}}
\newcommand{\la}{\lambda}
\newcommand{\ka}{\kappa}
\newcommand{\al}{\alpha}
\newcommand{\ga}{\gamma}
\newcommand{\de}{\delta}
\newcommand{\si}{\sigma}
\newcommand{\bomega}{\mbox{\boldmath $\omega$}}
\newcommand{\bsi}{\mbox{\boldmath $\sigma$}}
\newcommand{\bchi}{\mbox{\boldmath $\chi$}}
\newcommand{\bal}{\mbox{\boldmath $\alpha$}}
\newcommand{\bpsi}{\mbox{\boldmath $\psi$}}
\newcommand{\brho}{\mbox{\boldmath $\varrho$}}
\newcommand{\beps}{\mbox{\boldmath $\varepsilon$}}
\newcommand{\bxi}{\mbox{\boldmath $\xi$}}
\newcommand{\bbeta}{\mbox{\boldmath $\beta$}}
\newcommand{\ee}{\end{equation}}
\newcommand{\eea}{\end{eqnarray}}
\newcommand{\be}{\begin{equation}}
\newcommand{\bea}{\begin{eqnarray}}
\newcommand{\ii}{\mbox{i}}
\newcommand{\e}{\mbox{e}}
\newcommand{\pa}{\partial}
\newcommand{\Om}{\Omega}
\newcommand{\vep}{\varepsilon}
\newcommand{\bfph}{{\bf \phi}}
\newcommand{\lm}{\lambda}
\def\theequation{\arabic{equation}}
\renewcommand{\thefootnote}{\fnsymbol{footnote}}
\newcommand{\re}[1]{(\ref{#1})}
\newcommand{\R}{{\rm I \hspace{-0.52ex} R}}
\newcommand{\N}{{\sf N\hspace*{-1.0ex}\rule{0.15ex}%
{1.3ex}\hspace*{1.0ex}}}
\newcommand{\Q}{{\sf Q\hspace*{-1.1ex}\rule{0.15ex}%
{1.5ex}\hspace*{1.1ex}}}
\newcommand{\C}{{\sf C\hspace*{-0.9ex}\rule{0.15ex}%
{1.3ex}\hspace*{0.9ex}}}
\newcommand{\eins}{1\hspace{-0.56ex}{\rm I}}
\renewcommand{\thefootnote}{\arabic{footnote}}

\maketitle


\bigskip

\begin{abstract}
Two families of $SO(2n)$ Higgs models in $2n$ dimensional spacetime are
presented. One family arises from the {\it dimensional reduction} of
higher dimensional Yang-Mills systems while the construction of the other
one is {\it ad hoc}, the $n=2$ member of each family coinciding with the
usual $SU(2)$ Yang-Mills--Higgs system without Higgs potential. All models
support BPS 'monopole' solutions. The 'dyons' of the
{\it dimensionally descended} models are also BPS, while the electrically
charged solutions of the {\it ad hoc} models are not BPS.
\end{abstract}
\medskip
\medskip

\section{Introduction}

Field theoretic solitons find much application in various physical models.
Most recent applications of these are in the context of extra
dimensional theories, e.g. {\it large extra dimensions} with or without
gravity, when gravity and a negative cosmological constant are included
in AdS/CFT correspondence, and also in various theories
employing D$p$-branes. The special case, in which the soliton in question
is BPS is particularly pertinent in this last application, a prominent
role being played by the BPS dyons of the Yang-Mills--Higgs (YMH) model. The
generalisations of the $SU(2)$ YMH model in $4$ spacetime dimensions, to
all even dimensions, is the objective of this work. A brief discussion
of some of their possible applications will be given in the Summary section.
Here, we proceed directly to construct the models and their solutions.

When the dimensionality of the space on which the soliton lives is higher
than two, then the gauge fields~\footnote{There are solitons of ungauged
models, e.g. sigma models or some higher dimensional generalisations of
the Goldstone model, but without introducing a gauge field there
are no known systems which support BPS solitions {\it except} in two
dimensions.} must necessarily be non Abelian. Thus in any theory in
which the number of extra dimensions is larger than two, the construction
of BPS solitons is a pertinent task. The present work does just this,
by constructing BPS 'monopoles' of non Abelian Higgs models in arbitrary
even dimensions.

By 'monopole' in even dimensional spacetime, we mean a static solution
to a  YMH which is topologically stable. A BPS such
'monopole' is one for which the Bogomol'nyi (topological) lower on the
'energy' is saturated.

We present two distinct families of  YMH  models in
all even spacetime dimensions. The first is a class of models descending
from Yang-Mills (YM) systems in higher dimensions. The second family is
constructed in an {\it ad hoc} manner, guided only by the requirement
that there be a topological lower on the 'energy', and, that it is
saturated. (These features are automatically present
in the first class of models as a
result of the dimensional descent.) What is remarkable in the case of the
first class of models, namely those descended from higher dimensional YM,
is that the correspoding dyons are also BPS, directly generalising
this prominent property present in the familiar case of $d=4$ spacetime.

Subjecting a  YM  system in even Euclidean dimensions~\cite{T}
to dimensional reduction results in a residual YMH system. Depending on
the specific features of the dimensional descent, the residual YMH system
may\footnote{This is the case when the higher dimnsional space is a
product space whose extra dimension consists of
one compact coset space $K^N$ of $N$ dimensions, or, is the $N$ dimensional
torus $S^1\times S^1...\times S^1$, $N$ times.}~\cite{OT}, or may
not\footnote{This is the case when the extra dimension consists of the
product of more than one compact symmetric space, e.g.
$K_1^p\times K_2^q$, with $p\ge 2\ ,\ q\ge 2$, and $p+q=N$, in which case the
radius of compactification
of one of the two $K_{1,2}$ presents itself as a cosmological constant in
the residual model.}~\cite{cand}, inherit the topological lower
bound~\footnote{Assuming that the YM system in even dimensions is one
which is stabilised by the corresponding Chern--Pontryagin density.} of
the higher dimensional YM system. Restricting to the first type of
residual YMH systems, namely those supporting topologically stable
solitons, those descending from $4p$ Euclidean even dimensions have the
particularly simple property that they are characterised with {\bf only
one} dimensionful parameter, which is given by the 'radius' of
compactification, presenting itself as the Higgs VEV. We will henceforth
restrict to this simplest type of YMH models in our considerations of
the class of 'dimensionally descended' models.

In the present work, we consider
spherically symmetric BPS 'monopole' solutions of the two types of YMH
models just described. Concrete solutions will be constructed numerically.
For the YMH models descended from higher dimensional YM, the correspoding
'dyons' will also be given. In Section {\bf 2} we present the models,
subject them to spherical symmetry, and display their topological lower
bounds and Bogomol'nyi equations. In Section {\bf 3} we give the numerical
solutions and in Section {\bf 4} we summarise the results and give a
brief discussion.

\section{The BPS models}
There are two main families of non Abelian YMH models which can support
selfdual solutions that saturate the Bogomol'nyi bound, those which
are descended from a higher dimensional YM system~\cite{T}, and those
which are constructed in an {\it ad hoc} manner. The two classes are given
in separate subsections below.

Both families however share a common feature in their definitions, namely
the gauge group and its representation as well as the multiplet structure
of the Higgs fields are the same and are determined only by the
dimensionality $d=2n$ of the spacetime. For this reason we state these
at the outset.

In $d=2n$ dimensional spacetime, the gauge connection $A_{\mu}$ will take
its values in the algebra of $SO(d)$. Since $d$ is even, there are two
chiral representations of $SO(d)$ and we will take $A_{\mu}$ to be in one
or the other of these chiral represenations. The elements of the algebra
are represented in terms of the gamma matrices $\Gamma_{\mu}$ in $d$
dimensions and the corresponding chiral operator $\Gamma_{d+1}$
\be
\label{sigma}
\Sigma_{\mu\nu}^{(\pm)}=
-\frac{1}{4}\left(\frac{1\pm\Gamma_{d+1}}{2}\right)
[\Gamma_{\mu} ,\Gamma_{\nu}]\quad ,\quad \Sigma_{\mu\nu}^{(\pm)}=
-\frac{1}{4}\Sigma^{(\pm)}_{[\mu}\,\Sigma^{(\pm)}_{\nu]}\,,
\ee
the the spacetime index $\mu$ running over $0,i$, with the spacelike
index $i$ running over $1,2,...,(d-1)$. The Higgs field will be taken to
consist of a real isovector multiplet $\phi_i$ which can be expressed in
terms of the spin matrices \re{sigma} as
\[
\Phi=\phi_i\,\Sigma_{i,d}^{(\pm)}\,.
\]
In the following we will be repeatedly making use of the following
spinor identities satisfied by the spin matrices \re{sigma} in $d=2(p+q)$
dimensions 
\be
\label{spinor}
\Sigma(2p)=\pm\left(^{\star}\Sigma(2q)\right)(2p)\,,
\ee
where the $2p$-form $^{\star}\Sigma(2q)$ is the Hodge dual of the
$2q$-form $\Sigma(2p)$, and the
$\pm$ sign in \re{spinor} corresponds to the sign in \re{sigma}. The
$2p$-form spin matrix in \re{spinor} is the $p$-fold totally
antisymmetric product of the spin matrices \re{sigma},
\[
\Sigma(2p)=\Sigma_{\mu_1\mu_2...\mu_{2p}}
\]
and the Hodge dual of $\Sigma(2q)$ is
\[
\left(^{\star}\Sigma(2q)\right)(2p)=
\frac{1}{(2q)!}\vep_{\mu_1\mu_2...\mu_{2p}\nu_1\nu_2...\nu_{2q}}\,
\Sigma_{\nu_1\nu_2...\nu_{2q}}\,.
\]
Since the present study is restricted to {\it static spherically
symmetric} solutions, we state the fields $(A_{\mu},\Phi)$ subject to
this symmetry
\be
\label{YMHsph}
A_0=\eta\,a_0(r)\,\hat x_i\,\Sigma_{i,d}\,,\quad
A_i=\left(\frac{1-w(r)}{r}\right)\,\Sigma_{ij}^{(\pm)}\hat x_j\,,\quad
\Phi=\eta\,h(r)\,\hat x_i\,\Sigma_{i,d}\ ,
\ee
where $\eta$ is a constant with dimension of inverse length and the three
functions $(w,a_0,h)$ of $r=\sqrt{x_ix_i}$, are dimensionless. If
the model in question is descended from higher dimensions, then $\eta$
is the VEV of the Higgs field since in that case
the Higgs field has the same dimension as the connection. In
the other cases, where the Higgs field has other dimensions,
we will be modifying the third member of \re{YMHsph}.

It should be pointed out here that according to the static spherically
symmetric Ansatz \re{YMHsph}, all components of the YM
connection take their values in the $SO(d-1)$ subgroup of the {\it chiral}
representations of $SO(d)$. Accordingly, as a consequence of symmetry
breaking, the asymptotic gauge connections take their values in $SO(d-2)$.
This is most clearly seen in the Dirac gauge, in which there will be
a line singularity along the positive or negative $x_{d-1}$-axis,
where the asymptotic Higgs isovector points along the positive or negative
$x_{d-1}$-axis~\cite{OT}.

\subsection{Dimensionally descended models}
Higgs models arise from the dimensional reduction of YM models in higher
dimensions, and their various features depend on the particular mode of
dimensional descent, namely on the extra compact dimension and the
gauge group of the YM system in the higher dimension.

A remarkable feature of YMH systems descended from YM models in $4p$
Euclidean dimensions is that for a subclass of these models the
Bogomol'yi bounds can be saturated. This subclass of models consists
of those in $4p-1$ and $2$ Euclidean dimensions. The Bogomol'nyi
equations of all YMH models in the intermediate residual dimensions,
$4p-2$ down to $3$ are overedetermined~\cite{over} and are satisfied only
by trivial solutions, so they are excluded from consideration in this
work since we insist on BPS systems.
The class of models in $2$ Euclidean dimensions
are generalised Abelian Higgs models~\cite{BCT}, which do not interest us
here since we are concerned with higher dimensions and hence necessarily
non Abelian systems. Thus the family of $4p-1$ dimensional YMH models
just described, will be the main focus of our attention. For completeness
however, we will also consider all other YMH models in even dimensional
spacetimes (in odd dimensional Euclidean spaces) which do not result
from the dimensional reduction of higher dimensional YM systems but were
constructed in an {\it ad hoc} manner in \cite{1980}. Recently, the
spherically symmetric solution to the Bogomol'nyi equations of the $6$
dimensional example of these {\it ad hoc} models~\cite{1980}, was
constructed numerically by Kihara {\it et al}~\cite{6}.

This is the case of primary interest in this paper, namely the family of
models descended from the $p$-th YM system on $\R_{4p-1}\times S^1$ with
action
\[
\int_{\R_{4p-1}\times S^1}\,{\cal S}=
\int_{\R_{4p-1}\times S^1}\mbox{Tr}\,F(2p)^2=
\int_{\R_{4p-1}\times S^1}\mbox{Tr}\,F_{M_1M_2...M_{2p}}^2\,,
\]
$M_1=1,2,...,4p$, etc, which after integration over the coordinate on
$S^1$ yields the residual Higgs model on $\R_{4p-1}$
\bea
{\cal S}_{residual}&=&\mbox{Tr}\,\left[F(2p)^2+2p\,
F(2p-2)\wedge D\Phi\right]\nonumber\\
&=&\mbox{Tr}\,\left[(F_{m_1m_2...m_{2p}})^2+2p\,
F_{[m_1m_2...m_{2p-2}}\,D_{i_{2p-1}]}\Phi\right]\label{4pres}
\eea
$m_1=1,2,...,(2p-1)$, etc, and the square brackets denoting total
antisymmetrisation of the indices $m$.

Since the dimensional descent is by one step only~\footnote{If the descent
was over a larger number of dimensions, e.g. if $S^1$ were replaced by
$S^N$, $(N > 1)$, fixed YM curvature on $S^N$ would result in a symmetry
breaking leading to a smaller residual gauge group.} the gauge group of
the higher dimensional model is not broken as a result of the imposition
of symmetry effecting the descent, so it is the same in the
residual theory \re{4pres}. It is also clear that the Higgs field in
\re{4pres} has the dimensions of inverse length and $\eta$ is its VEV.

${\cal S}_{residual}$ in \re{4pres} is a YMH action density defined on a
Euclidean space of odd dimensionality $4p-1$. To generate a theory in
$4p$ dimensional (flat) Minkowski space, we introduce the time coordinate
$x_0$ by hand, such that the new coordinates are $x_{\mu}=(x_0,x_i)$, the
spacelike coordinates $x_i$ replacing $x_m$ in \re{4pres}. The
Lagrangian of the said model is now defined as
\bea
{\cal L}_{4p}&=&\mbox{Tr}\,\left[\frac{1}{2\,(2p)!}F(2p)^2
-\frac{1}{2\,(2p-1)!}F(2p-2)\wedge D\Phi\right]\nonumber\\
&=&\mbox{Tr}\,\left[\frac{1}{2\,(2p)!}(F_{\mu_1\mu_2...\mu_{2p}})^2-
\frac{1}{2\,(2p-1)!}
F_{[\mu_1\mu_2...\mu_{2p-2}}\,D_{\mu_{2p-1}]}\Phi\right]\label{4pmink}
\eea
the spacetime index $\mu_1$, etc, running over $0,i_1$, etc, with
$i_1=1,2,...,4p-1$, etc. Note that for $p=1$
\re{4pmink} is just the Lagrangian of the usual four dimensional YMH model.

The curvature $2$-form $F(2)=F_{\mu\nu}$ and the covariant derivative
$D_{\mu}\Phi$
\bea
F_{\mu\nu}&=&\pa_{\mu}A_{\nu}-\pa_{\nu}A_{\mu}+[A_{\mu},A_{\nu}]
\nonumber\\
D_{\mu}\Phi&=&\pa_{\mu}\Phi+[A_{\mu},\Phi]\nonumber
\eea
both take their values in the algebra of the gauge group, but not
the higher order forms appearing in \re{4pmink}. This means that
the definition of the model depends not only on the gauge group but
also on its representation, given by \re{sigma}.

\subsubsection{$A_0=0$: purely magnetic field}
In this case the static Hamiltonian pertaining to \re{4pmink} is
\bea
{\cal H}_{4p}&=&\mbox{Tr}\,\left[\frac{1}{2\,(2p)!}F(2p)^2+
\frac{1}{2\,(2p-1)!}F(2p-2)\wedge D\Phi\right]\nonumber\\
&=&\mbox{Tr}\,\left[\frac{1}{2\,(2p)!}(F_{i_1i_2...i_{2p}})^2+
\frac{1}{2\,(2p-1)!}
F_{[i_1i_2...i_{2p-2}}\,D_{i_{2p-1}]}\Phi\right]\label{4pstat}
\eea
which is bounded from below by
\be
\label{lbdesc}
{\cal H}_{4p}\ge\vep_{i_1i_2...i_{2p-3}i_{2p-2}i_{2p-1}}
\mbox{Tr}\,F_{i_1i_2}...F_{i_{2p-3}i_{2p-2}} D_{i_{2p-1}}\Phi\,,
\ee
the right hand side of which is a total divergence, by virtue of the
Bianchi identities, of the 'magnetic field'
\bea
{\bf B}&=&-\frac{1}{N_d}\mbox{Tr}\,
\Phi\,F\wedge F\wedge ...\wedge F\quad ,
\quad p\ \ \ \rm{times}\nonumber\\
B_{i_1}&=&-\frac{1}{N_d}\vep_{i_1i_2...i_{2p-3}i_{2p-2}i_{2p-1}}
\mbox{Tr}\,\Phi\,F_{i_2i_3}...F_{i_{2p-2}i_{2p-1}}\,,\label{magnetic}
\eea
where $N_d$ is the angular volume in the $d-1$ dimensional Euclidean
subspace. The general definitions of 'magnetic fields' in arbitrary
dimensions given in \cite{tz1} include \re{magnetic}.

The inequality \re{lbdesc} is saturated by the Bogomol'nyi equations
\bea
F(2p)&=&\pm\,^{\star}(F(2p-2)\wedge D\Phi)\nonumber\\
F_{i_1i_2...i_{2p}}&=&\pm\,
\vep_{i_1i_2...i_{2p}j_1j_2...j_{2p-1}}\,F_{j_1j_2}F_{j_3j_4}...
F_{j_{2p-3}j_{2p-2}}D_{j_{2p-1}}\Phi\,.\label{self85}
\eea
Subjecting \re{self85} to spherical symmetry according to \re{YMHsph},
and making use of the spinor identities \re{spinor} (with $q=p$ in the
case at hand) results in the reduced Bogomol'nyi equations
\bea
w'&\mp&\eta\,wh=0,\label{bogh85}\\
\eta\,r\left[(1-w^2)^{p-1}\,h\right]'&\pm&
\frac{(2p-1)}{r}\,(1-w^2)^p=0\,.
\label{bogw85}
\eea
The analytic proof of existence of the solution to \re{bogh85}-\re{bogw85},
for the case $p=2$, was given in \cite{yisong}.

Subjecting the static Hamiltonian \re{4pstat} to this symmetry
yields the one dimensional residual energy density functional
\bea
{\cal E}_{mon}&=&(1-w^2)^{2(p-1)}\left[2p\,w'^2+(2p-1)
\frac{(1-w^2)^2}{r^2}\right]\nonumber\\&&\,+\eta^2
\left[\frac{r^2}{2p-1}\left(\left[(1-w^2)^{p-1}\,h\right]'\right)^2
+2p(1-w^2)^{2(p-1)}\,w^2h^2\right]\,.\label{en1985}
\eea
It instructive to express \re{en1985} as
\bea
{\cal E}_{mon}&=&2p(1-w^2)^{2(p-1)}\left(w'
\mp\eta\,wh\right)^2\nonumber\\
&&+\left(\frac{\eta\,r}{2p-1}\left[(1-w^2)^{p-1}\,h\right]'\pm
\frac{(1-w^2)^p}{r}\right)^2\nonumber\\
&&\qquad\qquad\pm2\eta\,\frac{d}{dr}
\left[(1-w^2)^{2p-1}\,h\right]\,.\label{quad85}
\eea
Finally we state the asymptotic behaviours of the solutions
to \re{bogh85}-\re{bogw85},

\noindent
\underline{in $r\gg 1$ region}
\be
\label{1985gg}
h(r)\sim r^{-1}+o\left(r^{-3}\right)\quad,
\quad w(r)\sim\e^{-\eta r}
\ee
\noindent
\underline{in $r\ll 1$ region}
\be
\label{1985ll}
h(r)=2pb\,r^{2p-1}+o\left(r^{2p+1}\right)\quad,
\quad w(r)=1-br^{2p}+o\left(r^{2p+2}\right)\,.
\ee

\subsubsection{$A_0\neq{0}$: dyon fields}
As in the case of the dyons in $d=4$ spacetime~\cite{JZ}, we substitute
the full {\it static spherically symmetric }Ansatz \re{YMHsph} directly
into the Lagrangian \re{4pmink}. The consistency of this Ansatz can be
readily checked. The resulting static reduced one dimensional action
functional analogous to \re{en1985} now is
\bea
{\cal E}_{dyon}&=&(1-w^2)^{2(p-1)}\left[2p\,w'^2+(2p-1)
\frac{(1-w^2)^2}{r^2}\right]\nonumber\\&&\,+\eta^2
\left[\frac{r^2}{2p-1}\left(\left[(1-w^2)^{p-1}\,h\right]'\right)^2
+2p(1-w^2)^{2(p-1)}\,w^2h^2\right]\nonumber\\&&\,-\eta^2
\left[\frac{r^2}{2p-1}\left(\left[(1-w^2)^{p-1}\,a_0\right]'\right)^2
+2p(1-w^2)^{2(p-1)}\,w^2a_0^2\right]\,.\label{act1985}
\eea
It now follows from the form of the reduced action of the model in
$d=4p$ spacetime, just as for the familiar special case of
$p=1$~\cite{JZ}, that the following substitution
\be
\label{subst}
h(r)=f(r)\,\cosh\gamma\quad,\quad a_0(r)=f(r)\,\sinh\gamma\,,
\ee
with a constant parameter $\gamma$, renders the action functional
\re{act1985} identical to the energy functional \re{en1985}, with $h(r)$
in \re{en1985} now replaced by $f(r)$.

Since the second order Euler-Lagrange equations pertaining to \re{en1985}
are solved by the first order Bogomol'nyi equations
\re{bogh85}-\re{bogw85},
the solution $f_{sol}(r)$ of the latter then yields the self-dual dyon
solutions to \re{act1985} via the replacements \re{subst}.

In the absence of {\it electric--magnetic duality} in spacetimes of
dimensions $d>4$, in all cases with $p\ge 2$ the electric flux might be
defined as the flux of the following $1$-form field
\bea
{\bf E}&=&-\frac{1}{N_d}\mbox{Tr}\,A_0\,F\wedge F\wedge ...\wedge F\quad ,
\quad p\ \ \ \rm{times}\nonumber\\
E_{i_1}&=&-\frac{1}{N_d}\vep_{i_1i_2...i_{2p-3}i_{2p-2}i_{2p-1}}
\mbox{Tr}\,A_0\,F_{i_2i_3}...F_{i_{2p-2}i_{2p-1}}\,,\label{electric}
\eea
analogous to \re{magnetic}.

\subsection{The {\it ad hoc} models}
These models are defined in all even spacetime dimensions $d=2n$, and
are characterised by the Lagrangians
\be
{\cal L}_{2n}=\mbox{Tr}\,\left[\frac{1}{2\,(2p)!}F(2p)^2
-\frac{1}{2\,(2q-1)!}F(2q-2)\wedge D\Phi\right]
\label{2nmink}
\ee
analogous to \re{4pmink}. Note that for $p=q=1$
\re{4pmink} is just the Lagrangian of the usual YMH model.

Here, in \re{2nmink}, $q\neq{p}$ unlike in
\re{4pmink}. Here, our choices will include all possible $q$
within the range $1\le{q}\le(p-1)$, with $q=p$ omitted since that reverts
to the class of models already discussed in the previous subsection.
Unlike in the last class of models however, where $p$ is fixed by $4p=d$,
the choice of $p$ in \re{2nmink} is not restricted in that way. It is
nonetheless restricted by two other criteria. The first is the requirement
that there be first order Bogomol'nyi equations saturating the lower bound
of the static Hamiltonian (with $A_0=0$) pertaining to \re{2nmink}
\be
{\cal H}_{2n}=\mbox{Tr}\,\left[\frac{1}{2\,(2p)!}(F_{i_1i_2...i_{2p}})^2+
\frac{1}{2\,(2q-1)!}
F_{[i_1i_2...i_{2q-2}}\,D_{i_{2q-1}]}\Phi\right]\,,
\label{2nstat}
\ee
as a result of which $p$ and $q$ are restricted by $p+q=n$. The second
criterion is that the integral of the first term in \re{2nstat} be
convergent, i.e.
\[
\int_{\R_{d-1}}\mbox{Tr}F(2p)^2\sim\int\frac{r^{d-2}}{r^{4p}}dr=
\int\frac{dr}{r^{4p-d+2}}
\qquad,\qquad r\gg 1
\]
will be convergent only if
\be
\label{min}
4p\ge d=2n\,,
\ee
the minimum acceptable value of $p$ being given by $4p=d$, and the maximum
possible value being dictated by the antisymmetry of $F(2p)$, namely
$2p=d$.

Another difference of models \re{2nstat} from the dimensionally descended
models \re{4pstat} is that the Higgs field does not have the same
dimenisions as the connection.

Subject to these restrictions, there will numerous $SO(2n)$
Higgs models of the types \re{2nstat} supporting BPS 'monopoles' in
spacetime dimensions $d=2n$, their numbers increasing with $n$. Amongst
the plethora of such models, we will restrict our attention to a subclass
with $p=n-1$ and $q=1$, the $n=2$ case yielding the usual BPS monopole,
and the $n=2$ 'monopole' in $d=6$ constructed numerically recently in
\cite{6}. The Bogomol'nyi equations for these $(p=n-1,q=1)$ models are
\be
\label{self80}
F_{i_1i_2...i_{2p}}=\pm\,
\vep_{i_1i_2...i_{2p}j}\,D_{j}\Phi\,,
\ee
the $p=2$ member of which was proposed a long time ago in \cite{1980}. To
subject \re{self80} to spherical symmetry we employ the Ansatz
\re{YMHsph}, subject to a modification due to the fact that the Higgs
field in this class of models has the dimension of length raised to the
power of $2n-3$. We account for this by making the replacement
\[
\eta\longrightarrow\eta^{2n-3}
\]
in the third member of \re{YMHsph}, leaving the other two terms intact.
The resulting one dimensional Bogomol'nyi equations are
\bea
\frac{(1-w^2)^{n-2}}{r^{n-2}}\,w'&=&\mp\eta^{2(2n-3)}r^{n-2}\,w\,h,
\label{bogw80}\\
\eta^{2(2n-3)}r^{n-1}\,h'&=&\pm\frac{(1-w^2)^{n-1}}{r^{n-1}}
\label{bogh80}\,.
\eea
The reduced energy density functional corresponding to \re{2nstat} with
$(p=n-1,q=1)$ is
\bea
{\cal E}_{mon}&=&\frac{(1-w^2)^{2(n-2)}}{r^{2(n-2)}}\left[2(n-1)w'^2+
\frac{(1-w^2)^2}{r^2}\right]\nonumber\\
&+&\eta^{2(2n-3)}\left[r^{2(n-1)}h'^2+2(n-1)r^{2(n-2)}w^2h^2\right]
\label{en1980}
\eea
which can be rewritten as
\bea
{\cal E}_{mon}&=&2(n-1)\left[\frac{(1-w^2)^{n-2}\,w'}{r^{n-2}}
\pm\eta^{2n-3}r^{n-2}\,wh\right]^2+\left[\eta^{2(2n-3)}r^{n-1}\,h'\mp
\frac{(1-w^2)^{n-1}}{r^{n-1}}\right]^2\nonumber\\
&&\qquad\qquad\qquad\qquad\qquad\qquad\pm2\eta^{2(2n-3)}\frac{d}{dr}
\left[(1-w^2)^{n-1}\,h\right]\,,\label{quad80}
\eea
confirming \re{bogw80}-\re{bogh80}.

The solutions to \re{bogw80}-\re{bogh80} are the BPS 'monopoles' of the
class of models \re{2nstat} with $(p=n-1,q=1)$. Again, there will be dyon
solutions with $A_0\neq{0}$, but now these will not be given by the
BPS functions (evaluated numerically) via the substitution \re{subst}.
Rather, they will be solutions to the full second order Euler-Lagrange
equations. That these dyons are not BPS can be seen directly by
examining the reduced action density functional of the Lagrangian
\re{2nmink}, with $(p=n-1,q=1)$, subject to the Ansatz \re{YMHsph}
\bea
{\cal E}_{dyon}&=&\frac{(1-w^2)^{2(n-2)}}{r^{2(n-2)}}\left[2(n-1)w'^2+
\frac{(1-w^2)^2}{r^2}\right]\nonumber\\
&+&\eta^{2(2n-3)}\left[r^{2(n-1)}h'^2+2(n-1)r^{2(n-2)}w^2h^2\right]
\nonumber\\
&-&\eta^2\left[r^2\left(\left[(1-w^2)^{p-1}\,a_0\right]'\right)^2
+2(n-1)(2n-3)(1-w^2)^{2(p-1)}\,w^2a_0^2\right]\,\label{act1980}
\eea
which simply does not revert to the form of \re{en1980}, with $f$
replacing $h$, under the substitution \re{subst}.

The asymptotic behavious of the solutions to \re{bogw80}-\re{bogh80} are

\noindent
\underline{in $r\gg 1$ region}
\be
\label{1980gg}
h(r)\sim r^{-(2n-3)}+o\left(r^{-(2n-3+2)}\right)\quad,
\quad w(r)\sim\e^{-\frac{(\eta r)^{2n-3}}{2n-3}}
\ee
\noindent
\underline{in $r\ll 1$ region}
\be
\label{1980ll}
h(r)=(2b)^{n-1}\,r+o\left(r^{3}\right)\quad,
\quad w(r)=1-br^{2}+o\left(r^{4}\right)\,.
\ee

Before proceeding with the numerical construction, we examine briefly those
generic models \re{2nstat} which are
not subject to the restriction of $q=1$. In spacetime $d=10$ ($n=5$) there
is only one such model, characterised by $(p=3,q=2)$. In $d=12$ ($n=6$)
there is again only one such model, characterised by $(p=4,q=2)$.
n $d=14$ ($n=7$) there are two of these, characterised by $(p=4,q=2)$
and $(p=3,q=2)$, etc., their numbers increasing with $d$. 

For the rest of this section we will
restrict our attention to only the simplest example, namely that
in spacetime $d=10$ ($n=5$) with $(p=3,q=2)$. The Higgs field in this
example has dimension of length raised to the 3rd power. The static,
one dimensional energy density functional, with $a_0=0$, is
\bea
{\cal E}_{mon}&=&3\,\frac{(1-w^2)^4}{r^2}\left[\,w'^2+
\frac{(1-w^2)^2}2{r^2}\right]\nonumber\\&&\,+\frac16
\eta^6r^2\left[r^2\left(\left[(1-w^2)\,h\right]'\right)^2
+18(1-w^2)^2\,w^2h^2\right]\label{en2004}
\eea
which can be rewritten as
\bea
{\cal E}_{mon}&=&3\,(1-w^2)^2\left(\frac{(1-w^2)}{r}w'
\pm\eta^3\,r\,wh\right)^2\nonumber\\
&&+\frac16\left(\eta^3\,r^2\left[(1-w^2)\,h\right]'\mp
3\,\frac{(1-w^2)^3}{r^2}\right)^2\nonumber\\
&&\qquad\qquad\pm\,\eta^3\,\frac{d}{dr}
\left[(1-w^2)\,h\right]\,.\label{quad2004}
\eea
yielding the Bogomol'nyi equations
\bea
\frac{(1-w^2)}{r}\,w'&=&\mp\eta^3\,r\,w\,h,
\label{bogw2004}\\
\eta^3\,r^2[(1-w^2)h]'&=&\pm\,3\,\frac{(1-w^2)^3}{r^2}\,.
\label{bogh2004}
\eea

The asymptotic behaviours of the solutions of \re{bogw2004}-\re{bogh2004}
are

\noindent
\underline{in $r\gg 1$ region}
\be
\label{2004gg}
h(r)\sim r^{-3}+o\left(r^{-5}\right)\quad,
\quad w(r)\sim\e^{-\frac{(\eta r)^3}{3}}
\ee
\noindent
\underline{in $r\ll 1$ region}
\be
\label{2004ll}
h(r)=4b^2\,r+o\left(r^{3}\right)\quad,
\quad w(r)=1-br^{2}+o\left(r^{4}\right)\,.
\ee
\section{Numerical results}
All the Bogomol'nyi equations, \re{bogw85}-\re{bogh85},
\re{bogw80}-\re{bogh80}, \re{bogw2004}-\re{bogh2004}, can be expressed
as coupled first order ODE's in the dimensionless variable $\rho=\eta r$.

Both the {\it dimensionally descended} and the {\it ad hoc} models
have solutions with the correct asymptotics
only when the second derivative of the gauge function $w$ evaluated
at the origin, $w''(0)\sim b$,  takes on a certain value, 
which is dimension and model dependent. 
For example for the {\it dimensionally descended} models we have
$b(p=1)=1/6,~b(p=2)\simeq 0.0552096$ and  $b(p=4)\simeq 0.0176154$.
For the solutions of {\it ad hoc} models we find $b(d=6)\simeq0.7228039$,
$b(d=10)\simeq0.7400929$ and $b(d=12)\simeq0.7640163$. The corresponding value
for the only {\it hybrid ad hoc} model studied numerically is
$b\simeq 0.4593994$.

In both classes of systems, the models in spacetime dimension $d=4$
coincide and are identical to the usual $SU(2)$ YMH model in the BPS
limit. The Bogomol'nyi equations for this case are the only ones which
can be integrated analytically in closed form~\cite{PS} 
\footnote{The appearance of these non rational values of $b$ in the expansion
at the origin suggests that analytically evaluated solutions of higher $p$
monopoles, if they exist, should be parametrised by a set of functions
different from ($w,~h$).} . For all $d\ge 6$,
the solutions to these first order equations are constructed numerically.
We follow the usual approach and, by using a standard ordinary
differential equation solver, we evaluate the initial conditions at
$r=10^{-6}$ for global tolerance $10^{-14}$, adjusting for fixed shooting
parameter and integrating  towards  $r\to\infty$.
\newpage
\setlength{\unitlength}{1cm}

\begin{picture}(18,7)
\centering
\put(2,0.0){\epsfig{file=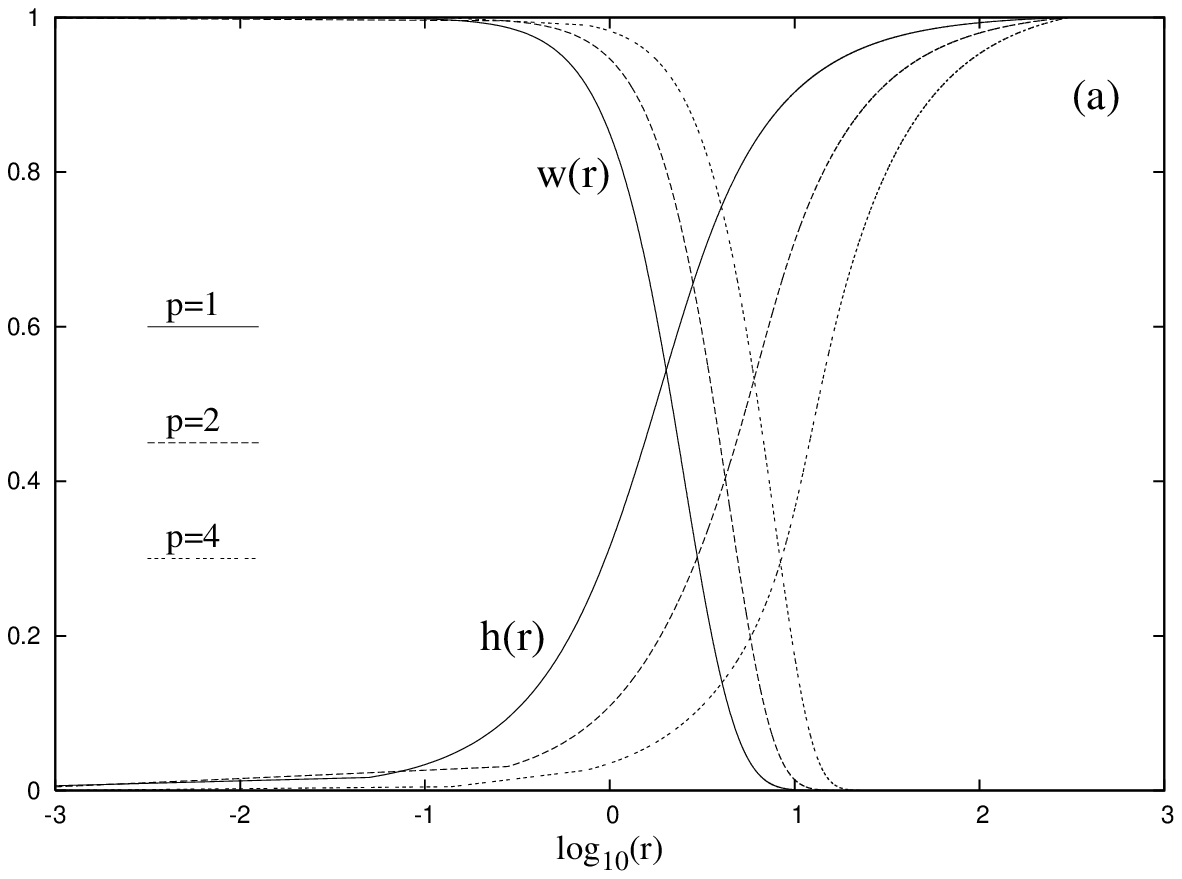,width=11cm}}
\end{picture}
\begin{picture}(19,8.)
\centering
\put(2.6,0.0){\epsfig{file=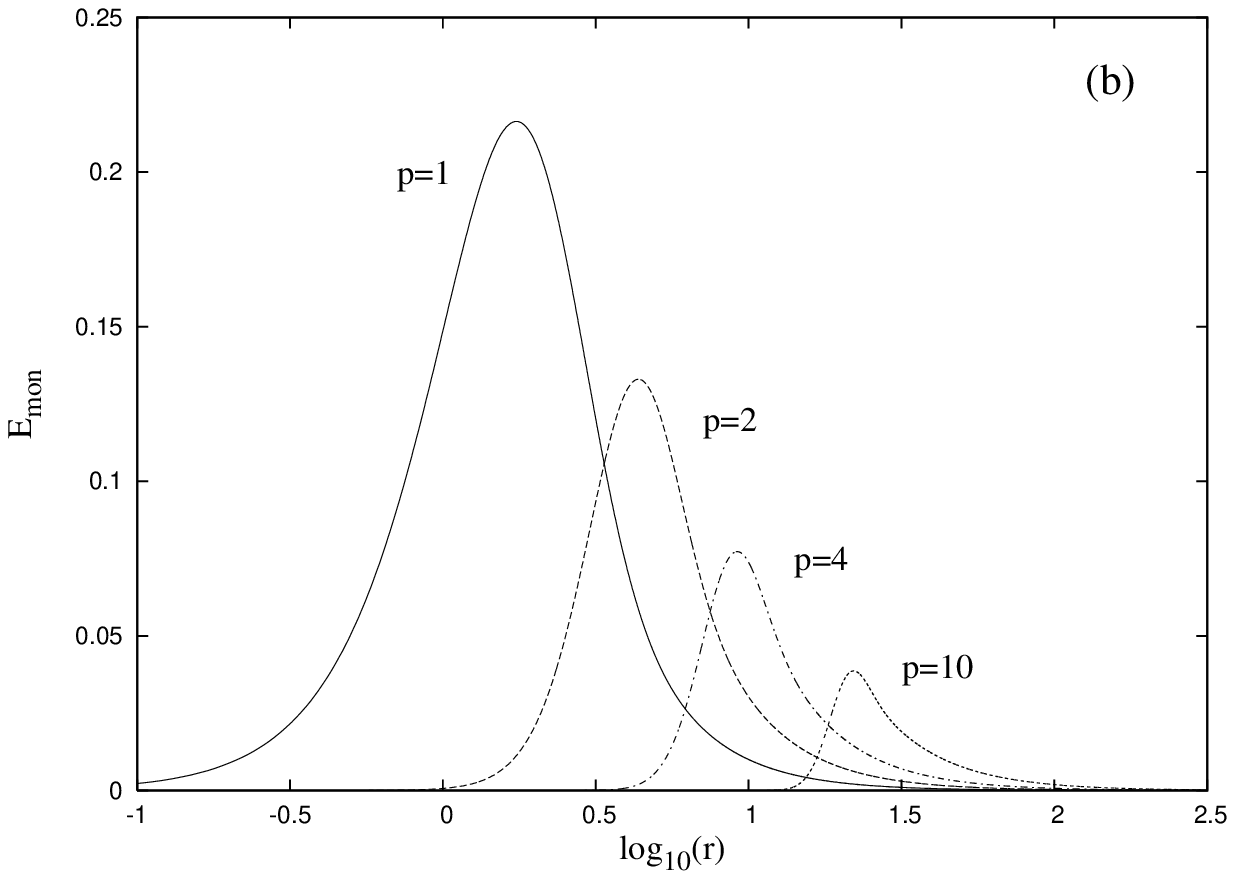,width=11cm}}
\end{picture}
\\
\\
{\small {\bf Figure 1.}
The profiles of the functions $w(r),~h(r)$ and the energy densities
${\cal E}_{mon}$ are shown for
several {\it dimensionally descended} models.}
\\
\\
In Figures 1a and 1b we present, respectively, the profiles of the
functions $w(r)$ and $h(r)$ corresponding to the BPS solutions of
equations \re{bogw85}-\re{bogh85} of the {\it dimensionally descended}
models, and their energy densities, for several values of $p$.
The same profiles for the BPS equations \re{bogw80}-\re{bogh80} and
energy densities are plotted in Figure 2a and 2b for the {\it ad hoc}
models. For completeness we present in Figure 3 the profile of the
solutions to equations \re{bogw2004}-\re{bogh2004} and its energy
density for the  ten dimensional monopole of the {\it hybrid ad hoc}
model with $(p=3,~q=2)$.

The qualitative properties of these solutions are the same as for the
well known $p=1$, $d=4$ BPS configurations~\cite{PS}.
The profiles of the functions $w(r)$ and $h(r)$ do not change
\newpage
\setlength{\unitlength}{1cm}

\begin{picture}(18,7)
\centering
\put(2,0.0){\epsfig{file=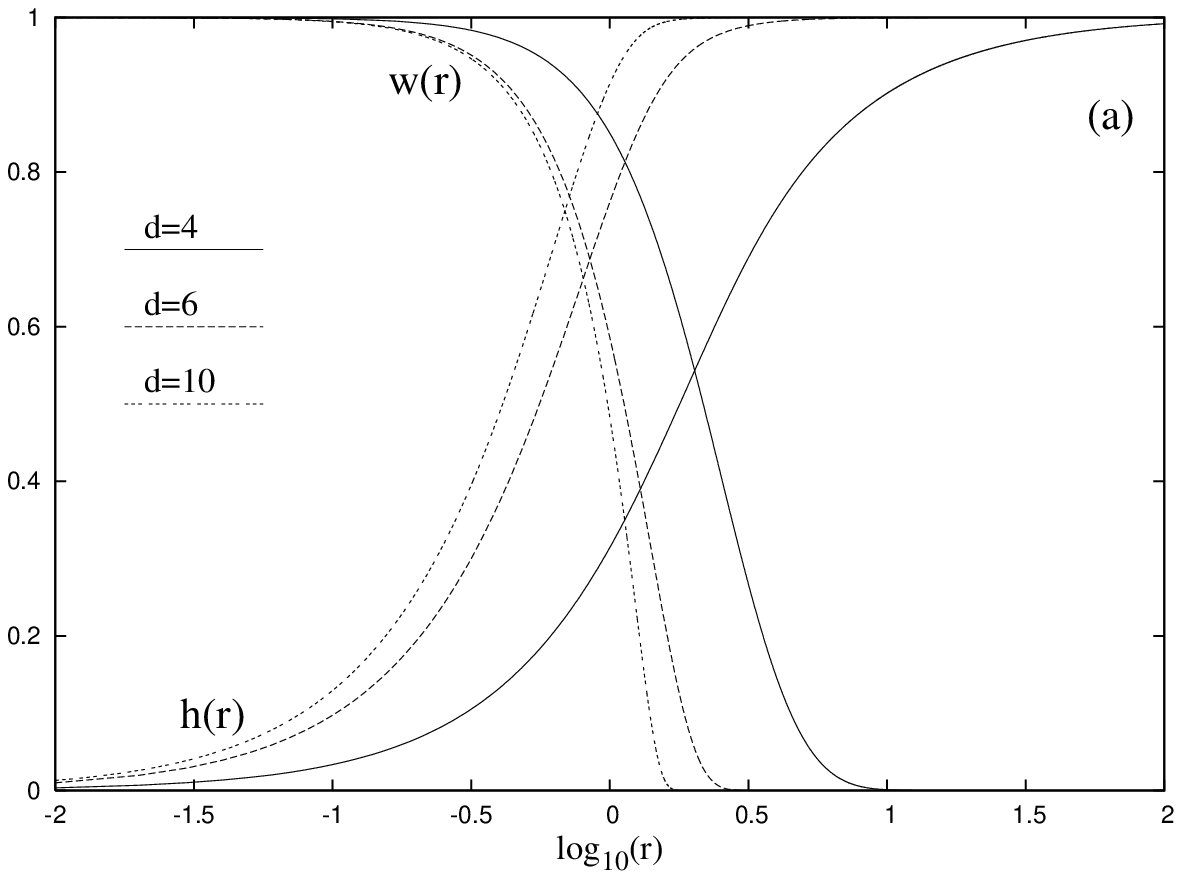,width=11cm}}
\end{picture}
\begin{picture}(19,8.)
\centering
\put(2.6,0.0){\epsfig{file=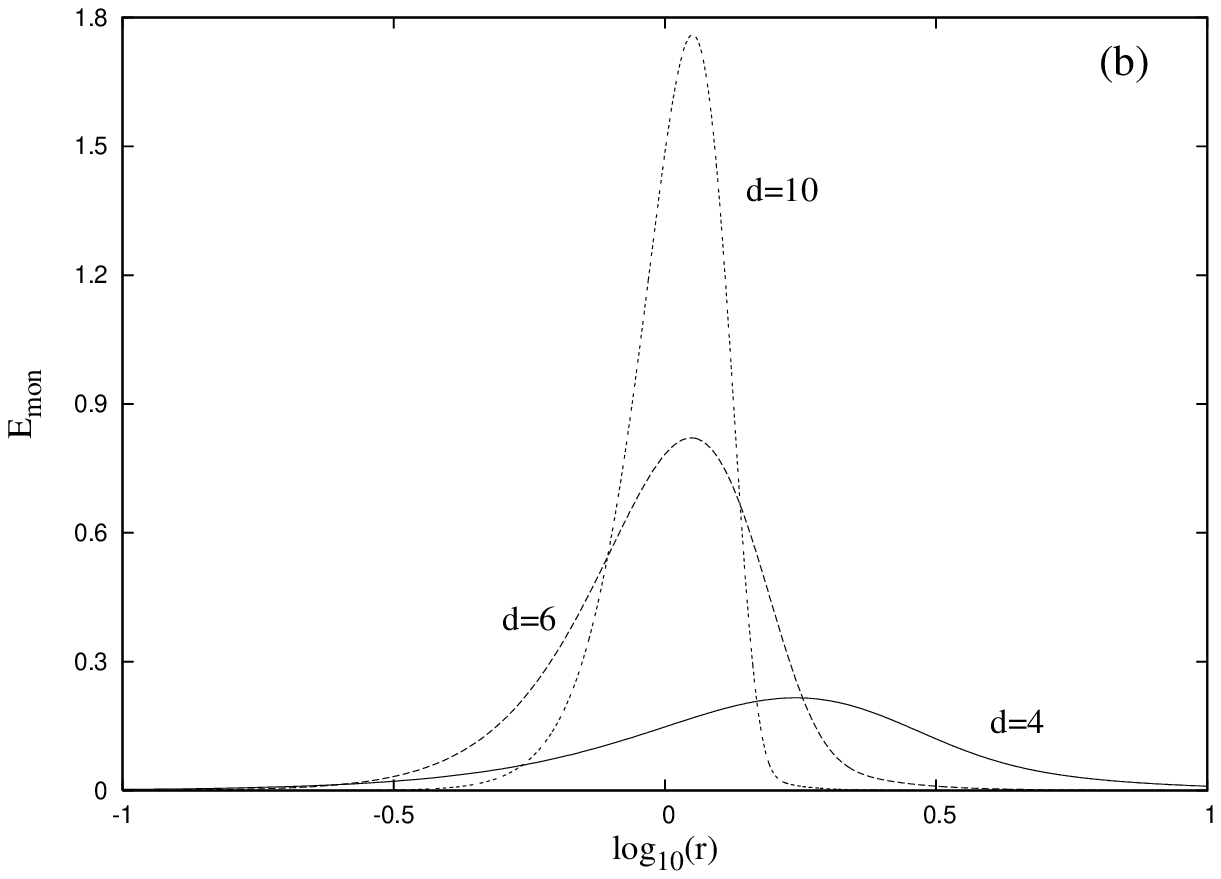,width=11cm}}
\end{picture}
\\
\\
{\small {\bf Figure 2.}
The profiles of the functions $w(r),~h(r)$ and the energy densities
${\cal E}_{mon}$ areshown for several {\it ad hoc} models.}
\\
\\
appreciably for the solutions living in different spacetime dimension.
The gauge and Higgs functions interpolate between the asymptotic values,
presenting no local extrema. The energies of these solutions are always
concentrated in a small region.

It turns out that in the case of the {\it dimensionally descended} models,
both the profiles of $w$ and $h$ as well as the peaks of the energy
densities move out from the origin with increasing dimension $d=4p$. Also,
the hights of those energy density peaks decrease with increasing $p$,
while the areas giving the total energies remain the same, with {\it unit}
normalisation due to spherical symmetry.

In the case of the sequence of {\it ad hoc} models by contrast, the
profiles of $w$ and $h$ move in 
\newpage
\setlength{\unitlength}{1cm}

\begin{picture}(18,7)
\centering
\put(2,0.0){\epsfig{file=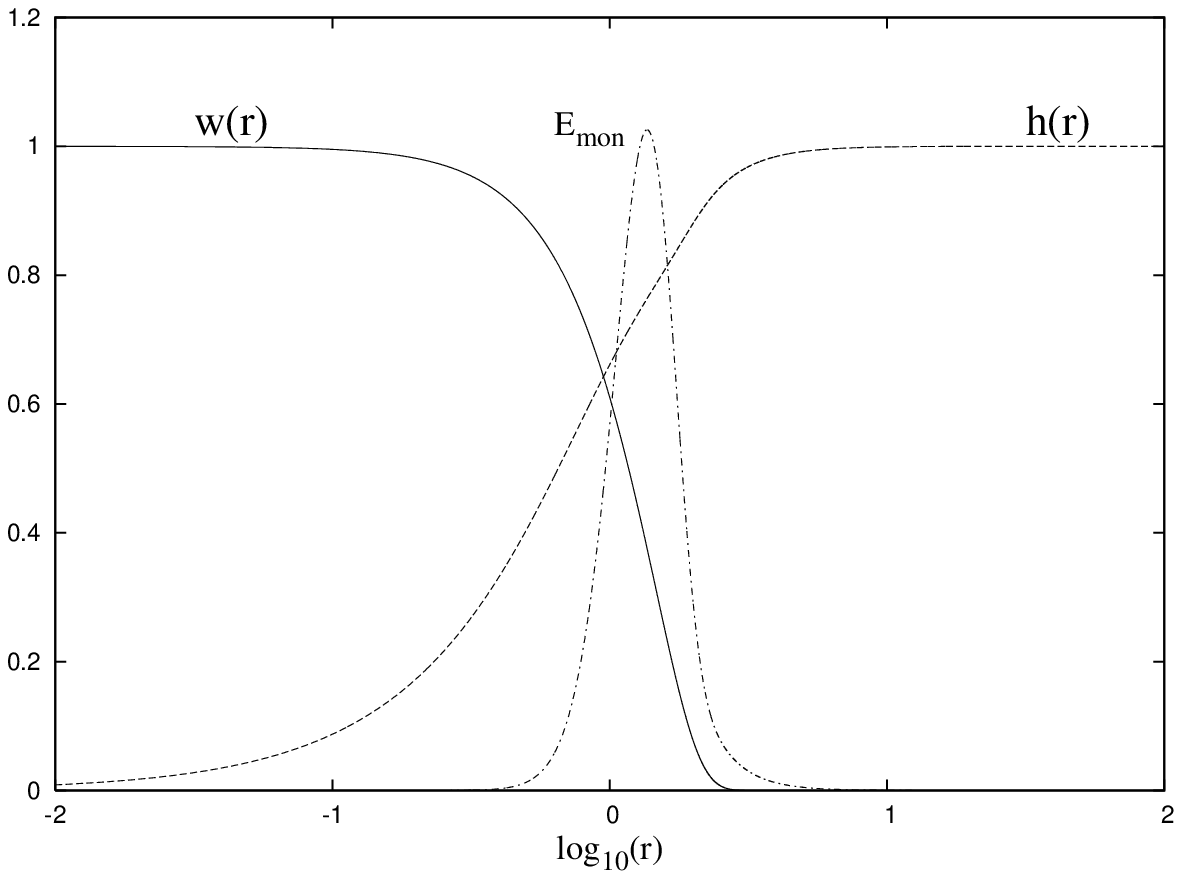,width=11cm}}
\end{picture}
\\
\\
{\small {\bf Figure 3.}
The profile of the functions $w(r),~h(r)$ and the energy density
${\cal E}_{mon}$ areshown for the ten dimensional {\it hybrid ad hoc} model.}
\\
\\
towards the origin with increasing $d=2n$.
The corresponding hights of the energy density peaks increase with
increasing $n$, again with {\it unit} area unchanged, and the positions
of these peaks seem to depend very weakly on on $n$.

\section{Summary and comments}
We have constructed topologically stable finite energy static solutions to
two families of Yang-Mills--Higgs models in all $d=2n$, even, spacetime
dimensions, subject to spherical symmetry in the appropriate dimension..

One of these families is arrived at via the dimensional
reduction of the $p$-th member of the Yang-Mills hierarchy on the Euclidean
space $d=(4p-1)\times S^1$ down to $4p-1$ Euclidean dimensions, whose solutions
then appear as the static solutions of the corresponding theory in $4p$
spacetime dimensions. These configurations are solutions of the appropriate
family of Bogomol'nyi equations, \re{self85} or \re{bogh85}-\re{bogw85}, and
hence saturate the respective topological lower bounds.
 
It may be interesting to note a particular feature of the Bogomol'nyi equations
\re{bogh85}-\re{bogw85}: For $p=1$, the solutions are found \cite{PS}
in closed form, while for $p=2$ an analytic proof for the existence of the
solution was given in \cite{yisong}. (This proof~\cite{yisong} can be adapted
to all $p$.) Substituting \re{bogh85} into  \re{bogw85} eliminates
the function $w$ and yieds a second order equation for $h$. 

For $p=1$, this
equation corresponds to the Liouville equation resulting from the selfduality
equations of $4$ dimensional $SU(2)$ YM subject to axial symmetry~\cite{W}.
Similarly, for $p\ge 2$, the second order equation in $h$ corresponds to the
generalisation of the Liouville equation resulting from the selfduality
equations of $4p$ dimensional $SO_{\pm}(4p)$ YM subject to axial symmetry,
presented in \cite{CST}. The analytic proof of existence to the latter
was given in \cite{spruck}.

The other class of models examined is arrived at in an {\it ad hoc} manner,
the only criterion being that the energies of the static solutions saturate the
topological (monopole) lower bounds by the appropriate Bogomol'nyi equations,
namely \re{self80} or \re{bogw80}-\re{bogh80}, and,
\re{bogw2004}-\re{bogh2004}. Typically, the Higgs fields in these models have
dimensions different from the inverse of a length.

The first member of both classes of models proposed, namely those in $d=4$
dimensional spacetime, coincide with the usual YMH model in the
Prasad--Sommerfield (PS) limit whose solutions are known in closed
form~\cite{PS}. The solutions of all the other models in higher spacetime
dimensions cannot be constructed in closed form and are evaluated numerically.

What is markedly more interesting about the first, dimensionally descended
class of models, is, that like the usual YMH model in the PS limit they also
admit dyon solutions obeying first order equations analogous to the
Julia--Zee dyons~\cite{JZ} in the PS limit. Of course
in spacetime $d\,>\,4$ the corresponding 'electric field', which we have
defined by \re{electric}, is not dual to the magnetic field \re{magnetic},
but is nonetheless there as a consequence of solutions with nonvanishing
electric potential $A_0$. This family of solutions generalising the Julia-Zee
dyons obeying first order equations is a feature only of the dimensionally
descended models introduced in section {\bf 2.1}, and not of the various
{\it ad hoc} models discussed in section {\bf 2.2}. In the latter case, there
are of course solutions with nonvanishing $A_0$, but these are subject to
the second order Euler--Lagrange equations rather than first order equations.

We conclude by making some brief remarks concerning the potential applicability
of the higher dimensional monopoles that we have presented above. For a start,
it is always of interest to see how the dimensionality of spacetime affects
the physical consequences of a given theory. Also, this type of objects might
form in the early universe when the present three spatial dimensions were not
yet separated from others, and a greater number of dimensions were equally
important.

The most immediate application, technically, is to proceed to the gravitating
case, thus enabling the study of the properties of gravitating monopoles in
higher dimensions, with reference to the detailed studies~\cite{BFM,Wein} in
$d=4$ spacetime.

Another direction to be explored straighforwardly is the construction of the
axially symmetric multimonopoles and dyons of the dimensionally descended
models. Since all our solutions obey Bogomol'nyi equations, all these solutions
are guaranteed to be topologically stable. Such dyonic solutions may be of
special interest in light of the lack of
electric--magnetic duality in higer dimensions.

Perhaps the most stringent test of our models is their status {\it vis \`a vis}
supersymmetry. Our Bogomol'nyi equations are first order, but they are
not linear in the curvature field strength (except in $d=4$ spacetime). This
contrasts with the BPS equations in $6$ and $8$ Euclidean dimensions, (not
involving Higgs fields) employed in \cite{BLP}. These are linear in the YM
curvature, unlike ours. On the other hand the energies of our models are
finite and bounded from below by topological charges. Should a way be found to
make our models respect supersymmetry, then they could be candidates for the
construction of field theory Supertubes~\cite{Tsd}, where Higgs fields feature.

\bigskip
\noindent
{\bf\large Acknowledgements} This work is carried out
in the framework of Enterprise--Ireland Basic Science Research Project
SC/2003/390 of Enterprise-Ireland.


\begin{small}

\end{small}

\end{document}